\documentclass{elsarticle}

\usepackage{amssymb}
\usepackage{amsmath}
\usepackage{natbib}
\setcitestyle{authoryear}
\setcitestyle{round}
\usepackage{graphicx}
\graphicspath{ {./images/} }
\usepackage{array}
\usepackage{url}
\usepackage[T1]{fontenc}
\usepackage[utf8]{inputenc}

\usepackage{xcolor}
\usepackage{soul}
\usepackage{lineno}
\usepackage{float}
\usepackage[top=1in, bottom=1in, left=1in, right=1in]{geometry}
\providecommand{\tightlist}{%
  \setlength{\itemsep}{0pt}\setlength{\parskip}{0pt}}

\newcommand{\expp}[1]{\mathrm{exp}(#1)}

\journal{Transportation Research Part A}

\begin{document}
\begin{frontmatter}

%% Title, authors and addresses

%% use the tnoteref command within \title for footnotes;
%% use the tnotetext command for theassociated footnote;
%% use the fnref command within \author or \affiliation for footnotes;
%% use the fntext command for theassociated footnote;
%% use the corref command within \author for corresponding author footnotes;
%% use the cortext command for theassociated footnote;
%% use the ead command for the email address,
%% and the form \ead[url] for the home page:
%% \title{Title\tnoteref{label1}}
%% \tnotetext[label1]{}
%% \author{Name\corref{cor1}\fnref{label2}}
%% \ead{email address}
%% \ead[url]{home page}
%% \fntext[label2]{}
%% \cortext[cor1]{}
%% \affiliation{organization={},
%%             addressline={},
%%             city={},
%%             postcode={},
%%             state={},
%%             country={}}
%% \fntext[label3]{}

\title{SimFLEX: a methodology for comparative analysis of urban areas for implementing new on-demand feeder bus services.}

%% use optional labels to link authors explicitly to addresses:
%% \author[label1,label2]{}
%% \affiliation[label1]{organization={},
%%             addressline={},
%%             city={},
%%             postcode={},
%%             state={},
%%             country={}}
%%
%% \affiliation[label2]{organization={},
%%             addressline={},
%%             city={},
%%             postcode={},
%%             state={},
%%             country={}}

\author[1]{Hanna Vasiutina\corref{cor1}}
\author[1]{Olha Shulika} 
\author[1,2]{Michał Bujak}
\author[2]{Farnoud Ghasemi} 
\author[1]{Rafał Kucharski}
\cortext[cor1]{Corresponding author: hanna.vasiutina@uj.edu.pl}

%% Author affiliation
\affiliation [1] {organization={Faculty of Mathematics and Computer Science, Jagiellonian University},
    addressline={Profesora Stanisława Łojasiewicza 6}, 
    city={Krakow},
    % citysep={}, % Uncomment if no comma needed between city and postcode
    postcode={30-348}, 
    % state={},
    country={Poland}}
\affiliation [2] {organization={Doctoral School of Exact and Natural Sciences, Jagiellonian University},
    addressline={Profesora Stanisława Łojasiewicza 11}, 
    city={Krakow},
    % citysep={}, % Uncomment if no comma needed between city and postcode
    postcode={30-348}, 
    % state={},
    country={Poland}}

%% Abstract -> 333 words

\begin{abstract}
On-demand feeder bus services present an innovative solution to urban mobility challenges, yet their success depends on thorough assessment and strategic planning. Despite their potential, a comprehensive framework for evaluating feasibility and identifying suitable service areas remains underdeveloped.

%METHOD
\textbf{Sim}ulation Framework for \textbf{F}eeder \textbf{L}ocation \textbf{E}valuation (SimFLEX) utilizes spatial, demographic, and transportation-specific data to run microsimulations and compute various key performance indicators (KPIs), including service attractiveness, waiting time reduction, and added value. Given the novelty of these services and the limited understanding of traveler behavior and preferences, SimFLEX employs multiple replications to estimate demand and mode choices and integrates OpenTripPlanner (OTP) for public transport routing and ExMAS (Exact Matching of Attractive Shared Rides) framework for calculating shared trip attributes and related KPIs. For each demand scenario, we simulate the traveler learning process using the method of successive averages (MSA) to stabilize the system. Following stabilization, we calculate the KPIs for subsequent comparative and sensitivity analyzes.

%EXPERIMENT AND RESULTS
We applied SimFLEX to compare two remote urban districts in Krakow, Poland - Bronowice and Skotniki - designated as candidates for the service launch.
Despite similar urban characteristics, our analysis revealed notable differences in main indicators between analyzed areas: Skotniki exhibited higher service attractiveness (approximately by 30\%) and added value (up to 7\%), whereas Bronowice showed greater potential for reducing waiting times (by nearly 77\%).
To assess the stability and reliability of our model output, we conducted a sensitivity analysis across a range of alternative-specific constants (ASC), assumed to be unknown. The results consistently confirmed that Skotniki is the superior candidate for feeder service implementation.

%SINGIFICANCE 
SimFLEX framework can be instrumental for policymakers to estimate new service performance in the considered area, publicly available and applicable to various use-cases. It can integrate alternative models and approaches, making it a versatile tool for policymakers and urban planners to enhance urban mobility.
\end{abstract}

%% Keywords
\end{frontmatter}

%% Add \usepackage{lineno} before \begin{document} and uncomment 
%% following line to enable line numbers
%% \linenumbers

%% main text
%%

\section{Introduction}
\label{sec:intro}

Thriving cities rely on efficient urban transport systems that ensure high accessibility, environmental sustainability, socio-economic progress, and public health \citep{nieuwenhuijsen2020urban}. The continuous expansion and growth of modern cities is driving the development of remote regions that offer residents access to affordable and high-quality living options. However, public transportation in these neighborhoods often suffers from low frequency and unreliability, resulting in poor accessibility for residents (\cite{ingvardson2018urban}, \cite{truden2022gis}). Consequently, residents of remote areas are often forced to depend on private vehicles due to the absence of viable alternatives, leading to increased traffic congestion, environmental pollution, and a decline in overall livability for both, drivers and non-drivers \citep{carroll2021identifying}.

Shared mobility, a rapidly evolving sector, presents an opportunity to enhance the effectiveness and sustainability of urban transportation by reducing the reliance on private cars and complementing traditional public transit \citep{Nikitas:2024, Becker:2020, Guyader:2021}. Feeder bus services, a type of shared mobility, usually operate in less accessible or peripheral urban areas, providing first- and last-mile connections.
These services connect residents of such sparsely populated regions with major public transportation hubs, such as train or metro stations, bus and tram terminals, etc. Depending on various factors, including population density, demand levels, and the availability of other transportation options, different types of feeder services can be implemented.

Demand-responsive feeder services are preferable when demand originates primarily at the transit station and disperses to multiple destinations, while fixed-route fixed-schedule services are preferred when demand is homogeneously distributed between the station and surrounding areas \citep{Calabrò:2022}. More flexible than fixed-route buses, on-demand feeders offer a more convenient, time- and cost-effective travel experience by dynamically adjusting their routes and schedules in response to real-time passenger demand \citep{Capodici:2024, Zhen:2024, Liyanage:2024}.
The advancement of technologies, combined with the widespread availability of user-friendly mobile applications, contributes significantly to the successful implementation and positive public perception of on-demand feeder bus services. However, despite their potential benefits, effectively implementing these services remains challenging due to various unknown determinants affecting performance and adoption. These include demand variability, user behavior and preferences, network integration, service reliability, and socioeconomic factors. Addressing these uncertainties requires continuous monitoring, adaptive adjustments, and iterative optimization to enhance efficiency, cost-effectiveness, and user satisfaction.

Beyond operational complexities, integrating these services successfully is also hindered by regulatory and policy constraints \citep{shaheen2020mobility}. City administrations often struggle to accommodate flexible, technology-driven mobility solutions within existing urban environments, while ensuring equitable access remains a pressing concern. Rigid transportation policies, outdated regulations, and conflicts with existing transit services can delay or limit the deployment of on-demand feeders. Furthermore, without targeted policies and inclusive planning, disparities in mobility options may persist, restricting the benefits of on-demand transit for diverse population groups, particularly those in underserved or lower-income areas. Overcoming these challenges requires municipalities to develop customized solutions that balance innovation with accessibility, fostering integration with existing public transport networks.

One of the key difficulties in deploying these services is predicting travel behavior and mode choice when the user perception is unknown. In such cases, indirect and predictive methods must be used to estimate mode choice and assess service effectiveness. Utility-based models in transportation analysis evaluate travel mode choices by incorporating factors such as travel time, detours, operational schemes, and transfers. These models account for adaptive travel behavior, learning processes, and multimodal route combinations, enabling a more dynamic and realistic representation of user decision-making \citep{Leffler:2024, Krauss:2022}. Agent-based and micro-simulation models replicate individual traveler movements within a transport network, providing insights into service demand and performance. Discrete choice models predict how users select between travel modes based on observed travel behavior and socio-demographic factors. Additionally, machine learning approaches use mobile phone data, GPS tracking, and transport system analytics to analyze movement patterns and predict demand. A combination of these models typically yields a more comprehensive and accurate assessment of travel behavior and service feasibility \citep{Alonso-González:2021, Narayanan:2023, Nahmias-Biran:2021, Schweizer:2021}.

In 2024, nearly 400 new on-demand transit projects were launched worldwide, marking a record high \citep{foljanty2024ridepooling}. However, most of these initiatives remain in the pilot phase, underscoring the difficulties of transforming conceptual models into sustainable businesses. Many on-demand transit systems have faced challenges such as low user adoption, operational inefficiencies, and financial failure due to the lack of a robust methodology for evaluating feeder service performance based on area-specific characteristics. Without a comprehensive evaluation framework designed for feeder services planning, deployments risk inefficiency, low user satisfaction, and financial losses.
This highlights the necessity of evaluating the feasibility of novel mobility solutions before deployment, considering the specific characteristics of the areas where they are intended to operate, and predicting user perceptions in situations where the service is still in the planning stage and actual demand remains unknown.

\subsection{Overview of the SimFLEX methodology}
\label{subsec:intro_method_overview}

\begin{figure}[h]
	\centering
		\includegraphics[width=1.0\textwidth]{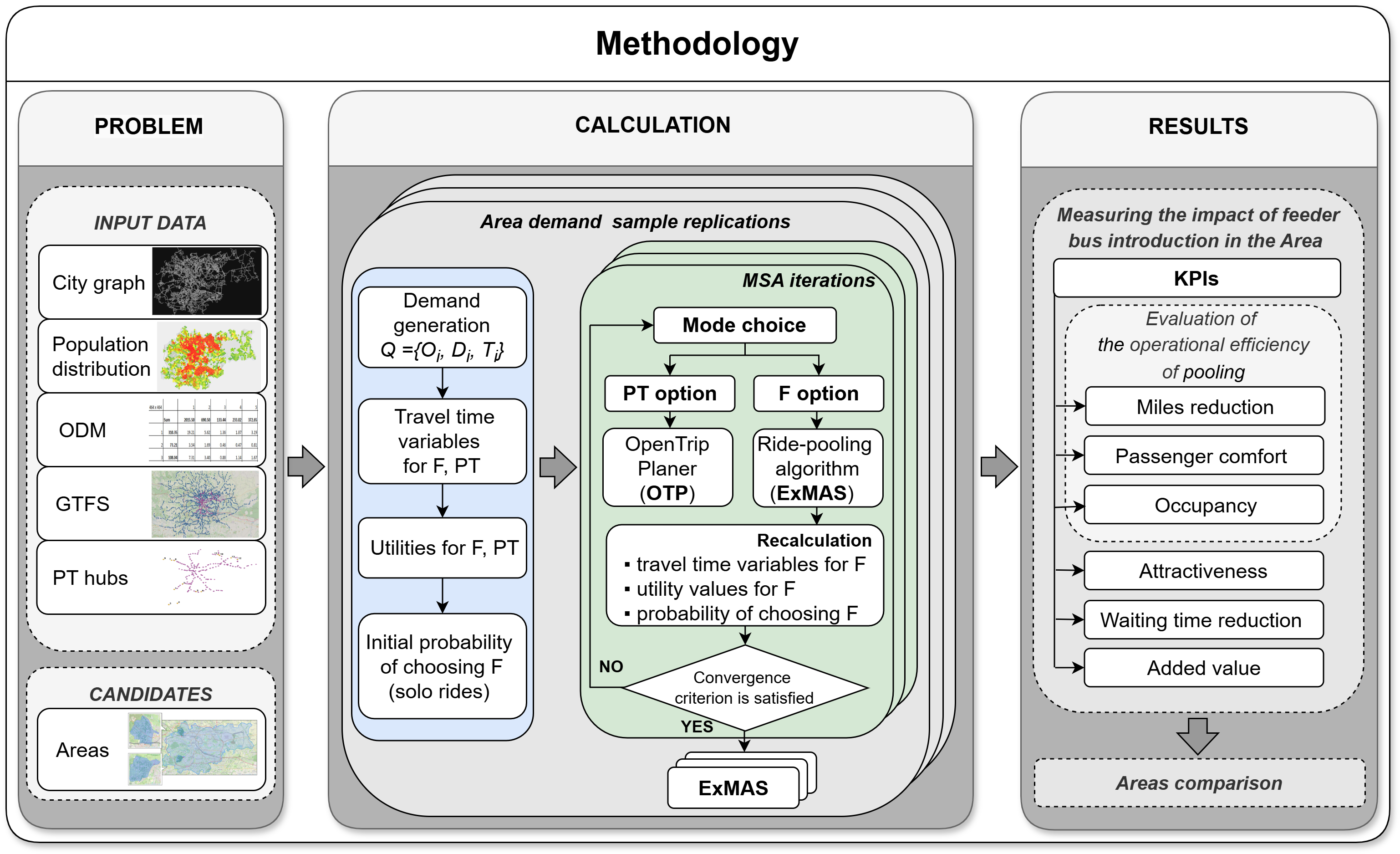}
	\caption{
SimFLEX computes service performance with the following methodology. For a given service area and hub location it uses widely available inputs (such as network graph, GTFS, population distribution and OD-matrices, detailed in \ref{subsec:method_workflow}) it runs a series of microsimulations to obtain a wide range of performance indicators. First, we sample microscopic demand patterns for services from macroscopic models (Section \ref{subsec:method_demand}). For each single demand realization, we simulate the travelers learning process (detailed in \ref{subsec:method_msa}), when they experience system performance (with unknown travel times due to detours, here sampled with ExMAS). After stabilization (when each travelers expectations meet the realizations) we simulate extra runs to compute indicators from the stabilized system (detailed in section \ref{subsec:method_outputs}). This concludes a single run of SimFLEX, which can then be replicated (for different realizations of the demand), or used for comparisons (between areas, hubs, parameterizations, etc.) as discussed in Section \ref{sec:results}.}
\label{fig:methodology}
\end{figure}

We address these challenges by introducing SimFLEX (\textbf{Sim}ulation Framework for \textbf{F}eeder \textbf{L}ocation \textbf{E}valuation), a methodology specifically designed to assess the feasibility and effectiveness of on-demand feeder bus services in diverse urban conditions. By leveraging spatial, socio-demographic, and transportation-specific data of the analyzed region, the method enables the computation of various KPIs for a given area-hub combination, allowing a comparative analysis to identify the most suitable urban area for service implementation (Fig.\ref{fig:methodology}). The proposed key indicators capture both operational aspects of the feeder buses, such as vehicle-hours traveled, passenger-hours, and vehicle occupancy, as well as utility-based metrics that reflect the effectiveness of the overall transport system that includes feeders as first- or last-mile solutions. These are service attractiveness, waiting time reduction, and overall added value.

Beyond evaluating feeder service effectiveness, SimFLEX is useful in comparing different urban areas, to introduce services where they offer the highest benefits. Additionally, it enables a sensitivity analysis of key performance indicators, notably those not well estimated (like alternative specific constant or transfer penalty). %which assesses the reliability of results under varying assumptions and model parameters.
To achieve these objectives, SimFLEX integrates a combination of computational tools, optimization techniques and analytical methods that together enable a comprehensive assessment of feeder system performance.
 
The methodology incorporates the following main tools and approaches:
\begin{enumerate}
\tightlist
    \item \textbf{Microscopic demand generation}. We use multiple replications to create diverse demand scenarios, capturing variability in traveler behavior. This involves disaggregating macro-level data for detailed individual-level analysis and applying a discrete choice model for mode choice modeling.
    \item \textbf{Effective and attractive matching for pooled rides}. The open-source ExMAS ride-pooling framework \citep{Kucharski:2020} is employed to compute shared trip attributes and feeder operational KPIs during learning and after system stabilization. It matches travelers into shared rides based on factors such as detour duration, waiting time, travel time savings, and cost reduction. ExMAS minimizes detours and enhances ride efficiency while considering service reliability and passenger comfort.
    \item \textbf{Day-to-day learning}, here implemented with MSA to simulate a process of learning variable system performance (feeder travel times), ensuring iterative system stabilization and convergence toward consistent results.
    \item \textbf{Public transport routing with OpenTripPlanner (OTP)}. The trip planning software OTP \citep{otp} is used to integrate feeder buses with public transport networks and to compare the performance of this integrated system with a public transport-only alternative. It serves as a public transport routing tool, incorporating different transport modes, such as buses, trains, trams, etc. OTP tool finds optimal routes based on real-time and scheduled transit data, considering travel time, number of transfers, and walking distances.
    % \item KPI sensitivity analysis. The reliability of the method outputs is assessed through a sensitivity analysis of KPIs under varying alternative-specific constants (ASCs) in the utility functions. Thereby, SimFLEX helps evaluate how changes in these unobserved factors affect the reliability and behavior of the model’s outputs.
    % \item Comparative area analysis. The computed KPIs allow for comparative analysis of different urban areas, helping to identify the most suitable locations for feeder service implementation.
\end{enumerate}

We propose SimFLEX as a comprehensive decision-support tool developed to address the lack of location-specific methods for evaluating the potential impact and feasibility of novel transportation services. The source code for the methodology is freely accessible on GitHub (Section \ref{sec:github}) and provides an adaptable foundation, enabling analysts to conduct analytically guided implementation decisions by simulating diverse scenarios, quantifying performance metrics, and analyzing potential outcomes.

\subsection{Literature review on demand-responsive services}
\label{subsec:intro_literature}

On-demand feeder bus services are constantly evolving, establishing themselves as a vital solution for last-mile passenger transportation \citep{Calabrò:2022}. Advances in ICT technology have facilitated the widespread adoption of these services, expanding access for a diverse range of users across various age demographics and socioeconomic backgrounds, especially those residing in areas with limited transportation access \citep{fleming2018social}.
By using real-time data, mobile applications, and automated dispatching systems, these demand-responsive services enhance convenience, affordability, and inclusivity \citep{khan2024new}. 
From a transport system perspective, the implementation of on-demand feeder services enhances operational efficiency by reducing travel time and costs while promoting sustainability through lower emissions and energy consumption related to transport \citep{naumov2023optimizing}. 
Modeling frameworks further support these potential benefits. \cite{Becker:2020}, using MATSim and a mode choice model,  demonstrate that shared mobility systems (car-sharing, bike-sharing, and ride-hailing) can decrease transportation energy use by 25\%.
Compared to traditional fixed-route buses, they can improve user satisfaction and efficiency, reducing mileage by up to 11\%, even despite a ridership reduction of up to 28\%  \citep{Coutinho:2020}. Additionally, the authors note the significant reductions in CO2 emissions - approximately by 45\%.
Utility-based models substantially contribute to understanding modal choices. \citet{diallo2025utility} apply a utility-based agent model, integrated with MATSim and multinomial logit model, to simulate intermodal travel behaviors in response to urban toll policies. Their experimental study demonstrates that a toll of \texteuro 20 (\textdollar 21.75) effectively reduces private vehicle usage by 20\%, encouraging a shift toward sustainable intermodal transport. These findings highlight the potential of policy-driven interventions to alleviate congestion and pollution by promoting public transit and park-and-ride facilities.

However, without prioritizing user satisfaction and thorough planning, on-demand feeder bus services risk failure.
Factors related to both traveler characteristics (such as age, car ownership, income, etc.) and service attributes (including travel time, cost, accessibility, and information availability) must be carefully considered \citep{Schasché:2022, Macfarlane:2021}. 
Similarly, the mode choice model developed by \citet{Narayanan:2023} shows that factors such as age, education, public transport ticket ownership, distance, and travel time significantly influence the use of shared mobility services (e.g., bike-, car-sharing, and ride-hailing).
Research on user preferences by \citet{Krauss:2022} further highlights that cost is a more significant factor than travel time in car-sharing and ride-pooling, whereas for other modes, such as private cars and public transport, both factors are equally important.
The analysis by \citet{Alonso-González:2021} indicates that the primary factor influencing ride-sharing decisions is the balance between time and cost, rather than the perceived drawbacks of sharing rides with others.
\citet{Geržinič:2023} state, that approximately 35\% of the analyzed sample demonstrated potential for adopting on-demand services, primarily encouraged by time-efficient (27\%) or cost-effective (9\%) travel option. Nevertheless, dedicated cyclists were unlikely to consider on-demand mobility a suitable option for urban journeys.

The role of microtransit services in improving user accessibility, especially in suburban areas, remains crucial. \citet{Capodici:2024} demonstrate this through simulation scenarios, conducted using the VISUM program, assuming an average 5-minute waiting time at stops. Their results indicate a potential increase in public transport ridership, with estimates suggesting a rise from 9\% to approximately 20\%.
A versatile analysis by \citet{Liyanage:2024} explores on-demand bus services from both, the traveler’s and the operator’s perspectives. For travelers, these services have the potential to enhance the commute experience by reducing average total waiting times by approximately 80–95\% during various peak periods throughout the day. From an operator's viewpoint, on-demand services demonstrate remarkable efficiency gains, with vehicle utilization rates improving by up to 70\% compared to traditional fixed-route bus services. Additionally, the research underscores a substantial reduction in transport-related emissions - approximately 50\% in CO2, 80\% in NO, and 40\% in PM10 pollutants.

Autonomous vehicles present a promising solution for on-demand transit services, attracting scientific interest due to their potential to transform urban mobility.
\citet{Hyland:2020} demonstrate that shared mobility systems, particularly autonomous vehicle fleets, can substantially improve the operational performance of the transport system, driven by demand density and network effects.
\citet{Fielbaum:2024} propose shifting part of the demand from fixed routes to flexible services by complementing fixed bus lines with a small fleet of on-demand vehicles operating alongside them. This scheme, with a few on-demand vehicles, can reduce the average walking time from 12 to 2 minutes while lowering overall costs by 13\%–39\% when using automated vehicles, and by over 10\% if vehicles remain non-automated.
Using the SimMobility simulator, \citet{Nahmias-Biran:2021} evaluate the impact of automated mobility on-demand services on the accessibility of different resident groups. The authors highlight that the city-wide deployment of such services increases accessibility and network performance, with greater benefits for low-income residents. 
Moreover, the use of automated shuttle bus services can lead to improvements in traffic flow and environmental conditions \citep{Oikonomoua:2023}. These findings are derived from traffic flow microsimulations conducted using the Aimsun Next software, by analyzing the impact of various operational schemes, considering different demand levels and operational speeds.
However, inappropriate or poorly designed regulations and policies regarding automated mobility may lead to the opposite consequences.
\citet{Oh:2020}, using agent-based simulation to analyze the impact of automated mobility-on-demand on the transportation system, warn that an unregulated implementation of such services could significantly increase traffic congestion and total travel distance.

Another effective strategy to realize the benefits of on-demand services implementation is the synchronization of the feeder schedules with main transit lines (such as metro, rail, or airport connection) to improve first- and last-mile connectivity. \citet{Gkiotsalitis:2022} estimates that such integration has the potential to reduce passengers' travel time by up to 10\%. 
Autonomous modular buses, offering greater flexibility for highly variable demand, could serve as feeder services to reduce operator and passenger costs \citep{Zermasli:2023}. The presented model uses a genetic algorithm to synchronize en-route feeder transfers, ensuring efficient connectivity between low-demand areas and multiple metro stations.
\citet{Calabro:2023} explore an Ant colony optimization algorithm within an agent-based model to design optimal feeder routes, demonstrating how new mobility modes can enhance public transport accessibility by maximizing efficiency, increasing profitability while minimizing travel times.

While studies highlight the potential benefits of demand-responsive services, particularly in enhancing first- and last-mile connectivity, real-world implementations provide evidence of their effectiveness in practice. Numerous feeder bus services have been successfully deployed, demonstrating how these models can be effectively integrated into existing transportation systems.

\subsubsection{Application of on-demand feeder services in real-world context}
\label{subsubsec:intro_feeder_examples}

Case studies illustrating the diverse outcomes and user perceptions associated with the implementation of demand-responsive transport systems in various urban conditions:
\begin{itemize}
\tightlist
    \item Mokumflex in Amsterdam, the Netherlands: the system received positive public feedback despite a decrease in ridership, achieving a significant reduction in kilometers traveled (\citet{Coutinho:2020}).
    \item Hüpfer in Vienna, Austria: \citet{Rossolov:2025} investigate the correlation between years of private car driving experience and willingness to adopt demand-responsive transport services. The findings reveal that approximately 91\% of survey participants believe that the longer individuals rely on private cars, the less likely they are to switch to demand-responsive transport. In contrast, only about 9\% perceive extended driving experience to have a positive influence on the adoption of such services.
    \item Breng flex in the Arnhem-Nijmegen region, the Netherlands: the results demonstrate that DRT improved generalized journey times for 50\% of the trips compared to fixed-route transit, with the most substantial reductions observed in areas poorly served by existing fixed-route options \citep{Alonso-González:2018}.
    \item BRIGJ in West Sydney, Australia: \citet{Perera:2020} state that such services can be integrated into the public transport network to fulfill several functions, such as: a peak feeder, complementing main transit lines during peak hours; a connector, enhancing accessibility and connectivity between areas, especially during off-peak hours; a coverage extender, serving areas with lower population density and limited access to traditional public transport.
    \item Guimabus in Guimarães, Portugal: the authors \citep{Martí:2025} highlight the potential of the service to reduce private vehicle use, particularly in peripheral areas, contributing to a more environmentally sustainable transportation system.
\end{itemize}

\subsection{SimFLEX contribution}
\label{subsec:intro_method_contribution}

Studies and successful case examples highlight the significant potential of on-demand feeder bus services to enhance urban mobility. With thorough evaluation and long-term planning, these services can effectively compete with private cars and complement public transport, improving the efficiency and sustainability of urban transportation networks. They not only increase user satisfaction by providing more convenient and flexible travel options, but also contribute to reducing energy consumption and transport emissions.

While these services offer potential, many examples highlight the challenges they face post-launch, often leading to failure. Common issues include financial losses for operators, declining ridership, and operational inefficiencies. Factors such as insufficient demand, poor integration with existing transport systems, and unsustainable business models contribute to these setbacks. These cases underscore the critical need for thorough pre-deployment analysis and focus on user satisfaction to ensure successful service implementation.

Simulation-based approaches and agent-based models are commonly employed to assess the efficiency of these services in varying operational scenarios. Utility-based mode choice models are widely used to include user perception, considering factors such as travel time, cost, and accessibility. However, these methods typically rely on data from questionnaires, pilot programs, and other sources, collected after the service has been launched, limiting their ability to guide pre-deployment decisions.

Despite the growing body of research, there remains a need for a comprehensive and systematic approach to evaluating the potential of on-demand feeder services, particularly when selecting areas for their implementation. While methods for assessing operational efficiency and user satisfaction are well-established, a research gap persists in existing approaches for measuring the impact of the service introduction and determining the proper locations for deploying these services, especially when the service is new and user perceptions have yet to be established.

The presented research aims to fill this gap by proposing a comprehensive methodology SimFLEX, which enables the evaluation of the potential of urban areas for implementing new on-demand feeder bus services, with a focus on service efficiency while enhancing user experience. 
% prioritizing user satisfaction.
Our methodology addresses the lack of systematic approaches available before the launch of new services. It integrates various models and techniques, including ExMAS for ride-pooling optimization, MSA for traveler learning and model stabilization, and OTP for service integration with public transport. By leveraging multiple microsimulations, the method allows for estimating travel demand in scenarios where only macro-level data is available. This approach enables the evaluation of service performance and user satisfaction without relying on prior knowledge of user perceptions.
By centering on user perception, SimFLEX helps identify favorable locations for successful service implementation, ensuring that the new service aligns with the needs and expectations of potential travelers.

This paper is organized as follows:
% Section 2 provides a literature review, 
Section \ref{sec:method} details the proposed methodology, including the models and techniques it incorporates. Section \ref{sec:results} demonstrates the practical application of the methodology through a case study. In Section \ref{sec:discussion}, we discuss our findings, summarize the main results, reflect on the limitations of our approach, and outline directions for future research.

\section{Methodology}
\label{sec:method}

SimFLEX is an iterative framework for assessing the effectiveness of on-demand feeder bus services. (Fig.\ref{fig:methodology}). It uses spatial and socio-demographic characteristics of the analyzed region, along with existing transportation system parameters, to estimate travel demand via a discrete choice model for mode selection. %The framework integrates the MSA method to stabilize the model and simulate the user learning process, employs the ExMAS framework to compute shared travel attributes and operational KPIs, and OpenTripPlanner for multimodal public transport trip planning. By calculating the system's key performance indicators, SimFLEX provides a basis for comparing areas potential and determining the most promising locations for implementing on-demand feeder bus services.
%Furthermore, by varying alternative-specific constants, the method enables sensitivity analysis of key metrics, allowing for an assessment of the model's stability and reliability, thereby providing a deeper understanding of whether observed outcomes reflect consistent patterns.

% In addition to these core inputs, the method incorporates various operational parameters, such as traffic conditions (e.g., peak traffic period), feeder bus characteristics (e.g., speed, capacity), travel time and cost attributes (e.g., value of time, willingness-to-share, delay), and other behavioral parameters.

SimFLEX is a two-loop framework: the outer loop generates multiple demand scenarios for analysis by sampling travel demand for feeder services, while the inner loop performs MSA iterations to model the traveler learning process and achieve system stabilization. 
First, this involves refining macroscopic data to generate detailed individual-level travel demand for agent-based simulations (Section \ref{subsec:method_demand}), which mimic the behavior and interactions of individual agents (here, travelers).
A discrete choice model allocates travelers among available transport modes based on utility functions that consider travel time, cost, and convenience factors (Section \ref{subsec:method_mode_choice}). The daily demand for the feeder bus is computed through this mode choice process, reflecting user preferences in response to service attributes and learning.
For each demand realization (i.e., a single sampled demand scenario), the inner loop simulates the traveler learning process, updating travel time expectations and mode choices until system stabilization is achieved.
% for the travel time variables, which are unknown due to detouring during the sharing trips.
Consequently, dependent on travel times, utilities mode choice probabilities are recalculated on each iteration of MSA for travelers selecting shared rides (Section \ref{subsec:method_msa}). The attributes of shared trips, including travel times, are obtained using the ExMAS framework.
For trips performed by public transport, the trip parameters - such as duration, walking distance, transit and waiting time - are obtained using the OTP software. 

Once the system stabilizes, we perform additional iterations to estimate key performance indicators.
In the outer loop, we perform multiple demand replications to capture variability in traveler behavior and network parameters.
These resulting metrics are then used to assess the effectiveness of introducing the feeder services in different urban areas and to conduct the comparative analysis (detailed in Section \ref{subsec:method_outputs}).

SimFLEX is designed to be adaptable, allowing researchers and practitioners to modify input data (such as varying transit schedules, population distributions, and operational constraints), assumptions, and models based on specific case study requirements. The modular framework enables integration with alternative demand estimation techniques, learning methods, and routing algorithms, making it applicable to various urban settings and transportation networks.
To ensure reproducibility, the complete SimFLEX implementation, including input data and computational functions, is publicly available in the GitHub repository (Section \ref{sec:github}). This allows for method validation, extensions, and further experimentation by researchers, urban planners, and policymakers interested in evaluating feeder service feasibility in different cities.

\subsection{Input data}
\label{subsec:method_workflow}

SimFLEX evaluates feeder service performance using a combination of spatial, demographic, and transportation system data. The primary inputs include:
\begin{itemize}
\tightlist
    \item city population distribution,
    \item public transport schedules in General Transit Feed Specification (GTFS) format,
    \item geographic boundaries of urban areas with designated hubs,
    \item origin-destination matrix (ODM) with spatially defined city zones,
    \item road network graph ($G$).
\end{itemize}

\subsection{Travel demand generation}
\label{subsec:method_demand}

Demand generation in SimFLEX creates a series of travel patterns while accounting for variations in individual traveler behavior and trip characteristics not covered in macroscopic demand models. To reflect demand variability and uncertainty, the framework generates multiple demand samples. It disaggregates macro-level data from the ODM (e.g., the number of trips during peak hours between city zones) into synthetic agents representing individual travelers.

Each replication \(N\) represents a distinct realization of travel demand resulting from macroscopic patterns and generated based on the population distribution within the analyzed area (\(P_A\)), the transportation network (\(G\)), and the movement patterns within the study region (ODM).
Specifically, each \(n\)-th demand realization is denoted as \(Q_n = f(\text{ODM}, G, P_A, T)\), where \(n \in [1, N]\) and \( T \) defines the temporal distribution of requests.
Resulting travel demand \(Q\) is a collection of individual requests:
\begin{equation}\label{eq:demand}
Q_n = \{\langle O_i, D_i, T_i \rangle \}.
\end{equation}
Each \(i\)-th traveler in SimFLEX has individual travel request, defined by an origin \(O_i\) (representing address points within the analyzed area), a destination \(D_i\) (various locations across the city), and a desired departure time \(T_i\) (randomly generated within a defined time range corresponding to the peak hour). 
Travel requests are mapped to the road network graph \(G = (V, E)\), representing the transportation network, where vertices (\(V\)) correspond to intersections and edges (\(E\)) denote road segments or transit connections. Each origin \(O_i\) and destination \(D_i\) is associated with the nearest vertex.%, and the travel paths are derived based on the graph structure.

\subsection{The mode choice model}
\label{subsec:method_mode_choice}

The SimFLEX methodology can be applied with various mode choice models and choice sets.
In this study, we consider a scenario in which travelers can choose between two available options:
\begin{enumerate}
\item
  \(F\) - an integrated trip, where an on-demand bus serves as a first- or last-mile connection from an origin to a designated hub. From the hub, public transport completes the trip to the final destination (the trip from the hub to the final destination is denoted as \(HD\)).
\item
  \(PT\) - a trip entirely serviced by the existing public transport system from the origin to the destination (denoted as \(OD\)).
\end{enumerate}

Therefore, we use the binary logit model to calculate the probability of the $i$-th traveler choosing the feeder service option \(F\) over the conventional public transport alternative \(PT\), we apply the following formula:
\begin{equation}
    P_{F_i} = \frac{\expp{\mu\cdot U_{F_i}}}{\expp{\mu\cdot U_{F_i}} + \expp{\mu\cdot U_{PT_i}^{OD}}},
    \label{eq:P_F}
\end{equation}
where \(U_{F_i}\) and \(U_{PT_i}^{OD}\) represent the utility functions for the integrated feeder bus and public transport options, respectively, and \(\mu\) is a scale parameter.

The initial expectation of \(U_F\) is based on an optimistic 'solo ride' scenario in which a traveler completes the journey independently, without sharing the ride with other passengers. This expectation is subsequently updated during the learning process.
% To determine initial mode choice probabilities, we establish a baseline scenario in which no ride-sharing occurs. 
% We refer to this solo ride scenario, where a traveler completes the journey independently, without sharing the ride with other passengers.

The utility of public transport for the $i$-th traveler, for both, the origin-destination \(OD\) and the hub-destination \(HD\) trip sections, is estimated by the total perceived cost of travel as follows \citep{Cats:2022}:
\begin{equation}\label{eq:utility_pt}
    U_{PT_i} = R_{PT} + \beta_t \cdot (\beta_{wk} \cdot t_{wk, i} + \beta_{wt} \cdot t_{wt, i} + \beta_{tr} \cdot n_{tr, i} + t_{transit, i}),
\end{equation}
where \(R_{PT}\) is a fare for public transport services, walking and waiting times (\(t_{wk, i}, t_{wt, i}\)), the number of transfers (\(n_{tr, i}\)), and the time spent onboard public transport (\(t_{transit, i}\)), from OTP . The relative significance of each factor in the decision-making process is captured by the \(\beta\) coefficients, which define the weight of each attribute in affecting the overall utility. Specifically, \(\beta_{wk}\), \(\beta_{wt}\) and \(\beta_{tr}\) represent the disutilities associated with walking, waiting, and transferring, respectively, reflecting their negative impact on the travel experience.
% The time-related components of the equation are amplified by walk and wait factor (\(\beta_{wk}\) and \(\beta_{wt}\), respectively) and the number of transfers is penalized by the transfer penalty coefficient (\(\beta_{tr}\)).

The utility of the feeder trip option \(F\) is calculated using the formula:
\begin{equation}
    U_{F_i} = U_{PT_i}^{HD} + \beta_t \cdot \beta_s \cdot (t^e_{trav_i} + \beta_w \cdot t_{w_i}) + ASC.
    \label{eq:U_F}
\end{equation}
Here, \(U_{PT_i}^{HD}\) is the utility associated with the public transport trip segment from the hub to the destination (calculated according to Eq. \ref{eq:utility_pt}). The behavioral coefficients - value-of-time (\(\beta_t\)), willingness-to-share (\(\beta_s\)), and delay sensitivity (\(\beta_w\)) - amplify the significance of time-related parameters: the expected travel time (\(t^e_{trav_i}\)) and the pick-up delay (\(t_{w_i}\)), computed using ExMAS.
%The ExMAS framework estimates travel times by considering the direct trip duration and additional delays caused by ride-pooling detours or waiting times.
The expected travel time \(t^e_{trav_i}\) is updated in each iteration of the MSA learning process (see Section \ref{subsec:method_msa}).
The alternative-specific constant \(ASC\) accounts for the average effect on the utility of all factors that are not explicitly included in the model, but still influence the choice of the feeder alternative.
Everyday of learning (iteration) the feeder demand is sampled based on the choice probabilities computed in Eq. \ref{eq:P_F} to form a set of travellers using the service on this particular day. %These selected travelers are subsequently used in the learning process during model stabilization.

\subsection{Model stabilization and learning via MSA}
\label{subsec:method_msa}

We apply the MSA technique to model the learning process of travelers who choose the feeder service and to stabilize the system as travel times and related parameters evolve over iterations. Specifically, this approach simulates how users learn and adapt their behavior over multiple days (iterations \(j\)), as they update their expectations based on their previous experiences, helping to smooth out fluctuations and generate stable travel time estimates.
This technique, a form of weighted averaging, can be substituted with alternative learning algorithms, such as exponential smoothing, Kalman filters or moving averages, to ensure convergence and account for dynamic changes in the system over time.

The process begins by setting an initial estimate \(t^e_{trav_{i,0}} = t_{trav_i}\), representing the expected travel time for an individual traveler \(i\) in a solo ride scenario. In each subsequent iteration \(j\), the expected travel time is updated as follows: 
\begin{equation}
    \begin{gathered}
        t^e_{trav_{i, j+1}} = (1 - \frac{1}{k_{i, j}}) \cdot t^e_{trav_{i, j}} + \frac{1}{k_{i, j}} \cdot t_{trav_{i, j}},
    \end{gathered}
    \label{eq:ttrav_e}
\end{equation}
where \(k_{i, j}\) is the number of trips that the traveler \(i\) has performed up to and including iteration \(j\), \(t^e_{trav_{i, j}}\) is the previous expected travel time, and \(t_{trav_{i, j}}\) is the actual travel time for the traveler \(i\), computed by ExMAS in iteration \(j\).

The MSA iteratively approximates expected travel times, utilities and mode choice probabilities.
In each iteration, the expected travel time \(t^e_{trav_i}\) in the utility function (Eq. \ref{eq:U_F}) is replaced with its updated value \(t^e_{trav_{i, j+1}}\), reflecting the traveler’s evolving expectations as a weighted average of past observations.
Over multiple iterations, this leads to a stable, convergent estimate of expected travel times, simulating a gradual learning process.

The learning procedure terminates when either a convergence criterion is satisfied or a predefined maximum number of iterations is reached. Here, we use the following convergence test:
\begin{equation}\label{eq:epsilon}
    \frac{|\bar{t}^e_{trav_{j+1}} - \bar{t}^e_{trav_{j}}|}{\bar{t}^e_{trav_{j}}} < \epsilon,
\end{equation}
where \(\bar{t}^e_{trav_{j+1}}\) and \(\bar{t}^e_{trav_{j}}\) denote the average expected travel times for the subset of travelers who selected the feeder service in iterations \(j\) and \(j+1\), respectively, and \(\epsilon\) is a predefined tolerance threshold.

% To ensure system stabilization, this criterion is checked over three consecutive iterations.

% By modeling daily travel decisions as an evolving process in which travelers refine their choices based on past experiences, MSA provides a realistic simulation of real-world traveler behavior, ultimately leading to stable travel time expectations and mode choices.

\subsection{Performance indicators}
\label{subsec:method_outputs}

After the system reaches a stable state, we perform additional iterations (days) to obtain statistically reliable performance metrics: operational KPIs for the feeder bus service and effectiveness indicators for the integrated feeder-public transport system. 
%We employ ExMAS to calculate operational KPIs that capture variations in service efficiency, user experience, and operator benefits.
Specifically, we derive the following indicators:
\begin{itemize}
\tightlist
    \item Decrease in total vehicle-hours (\(\Delta T_v\)): quantifies the change in efficiency for service operators by measuring the reduction in total vehicle operation time due to ride-pooling.
    \item Increase in total passenger-hours traveled (\(\Delta T_p\)): reflects the trade-off for travelers by indicating any increase in total travel time when shifting from solo rides to pooled services.
    \item Vehicle occupancy (\(O\)): assesses pooling effectiveness by comparing passenger hours with vehicle-hours, illustrating how efficiently shared rides utilize vehicle capacity. 
\end{itemize}
%These KPIs enable a comprehensive assessment of the feeder bus performance under different service scenarios \citep{Shulika:2024} and are used to compare the potential of urban areas for feeder service implementation.

To evaluate the overall effectiveness of the integrated feeder transit system and conduct a comparative analysis of urban areas, we incorporate the utilities for the feeder (\(U_F\)) and the direct public transport trip options (\(U_{PT}^{OD}\)), as well as the reduction in waiting time for public transport: for the feeder (\(t_{wt}^{HD}\)) and for the public transport trip option (\(t_{wt}^{OD}\)). 
Based on these SimFLEX outputs, we compute the following integrated system effectiveness indicators, expressed as average values over the \(n\)-th demand sample replication:

\begin{itemize}
\tightlist
    \item Attractiveness of the integrated feeder service: measuring the relative desirability of the travel options based on the utility differences:
    \begin{equation}
        \Delta A = \frac{1}{N} \sum_{n=1}^{N} (\bar{U}_{PT_n}^{OD} - \bar{U}_{F_n}).
        \label{eq:Delta_A}
    \end{equation}
    \item Reduction in waiting time for public transport: evaluates the effect of implementing feeder services on public transport waiting times:
    \begin{equation}
        \Delta T = \frac{1}{N} \sum_{n=1}^{N} (\bar{t}_{{wt}_n}^{OD} - \bar{t}_{{wt}_n}^{HD}),
        \label{eq:Delta_T}
    \end{equation}
    \item Overall added value: provides a quantitative measure of the improvement in traveler experience when using the new service \citep{Cats:2022}: 
    \begin{equation}
        \Delta V = \frac{1}{N} \sum_{n=1}^{N} \left( \ln(\exp(\bar{U}_{PT_n}^{OD}) + \exp(\bar{U}_{F_n})) - \bar{U}_{PT_n}^{OD} \right)
        \label{eq:Delta_V}
    \end{equation}
\end{itemize}

The proposed performance metrics are subsequently used for comparative analyses across different urban areas. This analysis helps identify locations where the introduction of feeder services would provide the greatest improvements in public transportation efficiency, accessibility, and overall user satisfaction.

%The framework enables the evaluation of diverse urban areas, transport hubs, and parameter configurations. By allowing decision-makers to assess multiple scenarios, SimFLEX supports the optimization of feeder service design based on the specific characteristics of each analyzed location.
%Moreover, SimFLEX supports a sensitivity analysis of the integrated system KPIs. Specifically, it examines how variations in key parameters, such as alternative-specific constants in the utility function, affect the overall performance of the integrated feeder service. 

The following section presents the results of the SimFLEX application, demonstrating the method's capabilities in a real-world scenario.

\section{SimFLEX application - a Krakow case study}
\label{sec:results}

% The SimFLEX methodology, presented in the previous section, is designed as a practical decision-support tool, enabling stakeholders to evaluate the effectiveness of planned feeder services before their implementation. The method offers a flexible solution applicable to diverse urban environments, allowing the integration of new functionalities and adjustments based on variations in input data. 

To demonstrate the effectiveness and practicality of SimFLEX, we apply it in a case study of two remote districts of Krakow, Poland - Bronowice and Skotniki (described in Section \ref{subsec:results_areas}). 
The analysis is structured as follows: first, we outline the experimental settings, including the input parameters and simulation configurations (Section \ref{subsec:results_config}) followed by the generation of demand based on Krakow-specific data (Section \ref{subsec:results_krk_demand}).  
We report the system stabilization and learning process (Section \ref{subsec:results_msa}), and feeder choice probability distribution (Section \ref{subsec:results_demand_distrib}). Following these foundational steps, we conduct a comparative analysis of two areas (Section \ref{subsec:results_area_comparison}) and chceck if the findigns remain invariant to behavioural parameter ($ASC$) in Section \ref{subsec:results_asc_analysis}.

% Specifically, the SimFLEX is used to assess the effectiveness of a planned on-demand feeder bus service in each area by calculating KPIs. By comparing the outputs across the two districts, we identify the area with the highest potential for service implementation. Finally, we conduct a sensitivity analysis to explore how variations in ASC values influence the method's outcomes.

\subsection{Characteristics of the study areas}
\label{subsec:results_areas}
 
Bronowice and Skotniki, two developing districts of Krakow, face significant transportation challenges, such as infrequent or nonexistent public transport services, poor infrastructure, and narrow, one-way streets that complicate mobility. 
To address these issues, Krakow's public transport authorities plan to introduce on-demand feeder buses in one of these low-density districts by winter 2025, requiring an analysis of each area's potential \citep{shulika2024selecting}.
The proposed system is set to operate via the mobile application that collects ride requests, utilizing two 6-seater buses to pick up residents from designated stops and transport them to nearby public transport hubs – major interchange points connecting to Krakow’s high-frequency tram and bus lines.
The main challenge for the authorities, however, is selecting the most suitable district for service implementation, especially given the uncertainty surrounding future user demand.

\begin{figure}[!h]
	\centering
		\includegraphics[height=7cm,keepaspectratio]{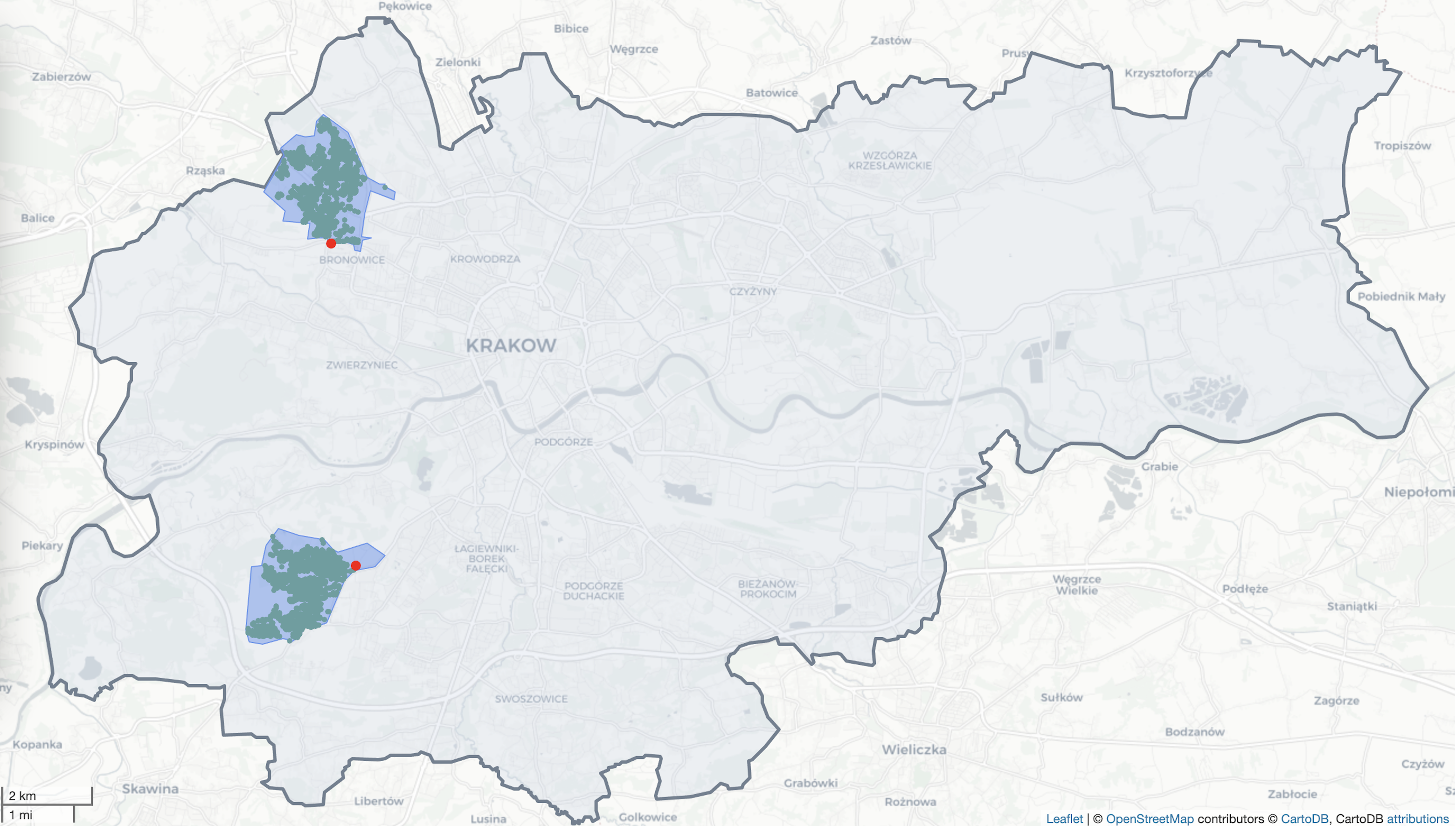}
	\caption{Two candidate areas in Krakow (Poland): Bronowice (top-left) and Skotniki (bottom-left) as candidate locations for on-demand feeder bus implementation, with address points (green dots) and public transport hubs (red circles) marked.\\
    The proposed feeder system will function on demand, allowing residents to request rides via a mobile application. Small 6-seater buses will pick them up from designated stops within their district and transport them to the respective public transport hub (Bronowice Małe for Bronowice, and Czerwone Maki P+R for Skotniki) for onward travel on Krakow's tram and bus lines.}
    \label{fig:map_areas}
\end{figure}

The districts preselected for assessment, Bronowice and Skotniki (Fig. \ref{fig:map_areas}), share similar characteristics, including their area (approximately 4.4 km2), population (8,390 in Bronowice and 6,070 in Skotniki), population distribution and transport network structure. The average distance from residential address points to transport hubs is about 1.4 km in Bronowice and 2 km in Skotniki. Major public transport hubs, Bronowice Małe and Czerwone Maki P+R, were selected to serve as the final stops for feeder buses near the Bronowice and Skotniki area, respectively. Both hubs offer direct access to multiple tram and bus connections to various destinations across the city.
% top - [t!]

\subsection{Experiment configuration}
\label{subsec:results_config}

The data for the experiment were collected from multiple sources. Krakow public transport authorities provided access to:
\begin{itemize}
\tightlist
    \item spatially defined urban areas, represented as polygons that specify the boundaries for planned feeder service implementation, with assigned public transport hubs locations,
    \item city population distribution dataset, providing information on locations of address points within the city,
    \item city zones, represented as multipolygons, 
    \item ODM that contains the expected number of trips between city zones.
\end{itemize}

The transport network and schedule data were obtained from open-access platforms, such as OpenStreetMaps (using the OSMnx python library) and \citet{MobilityDatabaseKrakow}:
% (\hl{https://gtfs.ztp.krakow.pl/}).
\begin{itemize}
\tightlist
    \item transit schedules in general transit feed specification (GTFS) format, providing detailed information on public transit routes, stops, and schedules,
    \item city road network graph.
\end{itemize}

Next, we define the simulation attributes, which reflect key parameters that influence the operation of the system and remain constant across the simulations
: %Each parameter plays a distinct role in the model, affecting traveler decisions and overall system behavior:
\begin{itemize}
\tightlist
    \item value of time ($\beta_t$): 43.3 PLN per hour (\ref{eq:U_F}),
    \item transfer penalty (\(\beta_{tr}\)): 500, representing the perceived inconvenience of transfers (\ref{eq:utility_pt}),
    \item walk and wait factors (\(\beta_{wk}\) and \(\beta_{wt}\)): 2, scaling travel times to reflect the perceived inconvenience of walking and waiting (\ref{eq:utility_pt}),
    \item feeder bus average speed: 6 m/sec,
    \item transfer time ($t_w$): 30 seconds, representing the delay between pick-up and drop-off operations,
    % \item public transport fare (\(R_{PT}\)): 0.175 \hl{units?},
    \item convergence criterion ($\epsilon$): 0.001 (\ref{eq:epsilon}).
\end{itemize}

To ensure a thorough analysis and statistical validity of the results, we run SimFLEX with the following setup (Fig. \ref{fig:methodology}):
\begin{itemize}
\tightlist
    \item number of demand replications (\(N\)): we perform 100 replications to account for the stochastic nature of demand patterns,
    \item MSA iterations (\(j\)): we simulate the learning process for 30 iterations (days) to allow the system to reach a stable state,
    \item post-stabilization iterations (\(j\)):  we conduct additional 20 iterations (days) after stabilization to obtain reliable KPI estimates.
\end{itemize}
For the sensitivity analysis, we use the following setup: number of demand replications is set to \(N\) =  5. For each demand replication, 50 ASC values were sampled evenly from \([0, 5]\).

We make the following assumptions for the experiment:
\begin{enumerate}
\tightlist
    \item The ODM used in the analysis is based on 2017 data. To account for population growth, we apply an adjustment coefficient, allowing to approximate the data to 2023 and assume the data remains applicable.
    \item Initially, we determine the ASC for the study areas under the solo ride option. 
    The ASC is calculated as the value that yields an expected probability of choosing the feeder service \(E(P_{F}) = 0.3\).
    \item Our analysis considers only two available transportation options: an integrated feeder (\(F\)) or a public transport trip (\(PT\)). However, the methodology is flexible and can be extended to incorporate additional transportation modes.
\end{enumerate}

\subsection{Sampling trip requests}
\label{subsec:results_krk_demand}

We simulate a 30-minute operational period of morning peak hour, capturing the highest travel demand from homes distributed across the area towards the hub.
We disaggregate macro-level demadn matrices to individual trip requests \(Q_i = {O_i, D_i, T_i}\), where each request is defined by an origin (\(O_i\)), a destination (\(D_i\)), and a departure time (\(T_i\)), originating within the study area during peak hour and terminating at various destinations across the city.

Since ODM data contains transport flows between urban zones, whose boundaries may not align with the study area's borders, we calculate the area's total trip production (sample size \(S_A\)) using a proportional relationship between a zone’s total trip production (\(S_{Z_i}\)) and its population (\(P_{Z_i}\)). We assume this same ratio applies to the population within the study area inside zone \(i\) (\(P_{A_i}\)), leading to the following calculation:

\begin{equation}
    S_A = \sum_{i \in Z_A} \left( S_{Z_i} \cdot \frac{P_{A_i}}{P_{Z_i}} \right),
\end{equation}
where \(Z_A\) represents the set of zones that intersect the study area.

To model trip distribution, we use destination probabilities derived from the ODM. For each randomly sampled traveler originating in the study area, we identify the corresponding ODM row representing the distribution of trips from their origin zone to all other destination zones. 
Then, using a probability-weighted selection based on each destination's relative share of the total outbound flow, we assign a destination zone. %After a destination zone is selected, its geographic centroid is used to determine the final destination coordinates for the simulated trip.
%This approach enables the disaggregation of Krakow’s macro-level ODM data into individual trip requests, producing a detailed and localized representation of travel demand within the study area.

We repeat this demand sampling process 100 times (as described in Section~\ref{subsec:results_config}), with 213 travelers for Bronowice and 155 for Skotniki in each replication. During each MSA iteration, a subset of these travelers is selected for feeder trip option, averaging 53 travelers in Bronowice and 41 in Skotniki.

% Following the demand estimation, we model the mode choice, determining which travelers select the feeder bus option. For these travelers, we then simulate the learning process to explore how users adapt to the new service over time.

\subsection{Analysis of system stabilization and traveler learning}
\label{subsec:results_msa}

%SimFLEX enables the analysis of traveler learning process, capturing how traveler choices and expected travel times evolve over time. By simulating iterative adjustments using the MSA, the method enables analysis into how users of the feeder bus system adapt their behavior over time.
%To investigate this process, we conduct a series of simulation runs, performing 30 iterations of the MSA for each of the 100 area demand sample replications, as defined in the experiment settings in Section \ref{subsec:results_config}). 
%This simulation helps to observe the convergence patterns and stability of travel times across a wide range of demand scenarios. 
We report the number of days (iterations) required for the system to converge. While individual iterations may vary, analysis across multiple runs for both Bronowice and Skotniki areas shows that convergence typically occurs around iteration 16. Therefore, we continue the simulations for the full 30 iterations, ensuring that outputs are estimated only after stabilization is achieved. Additionally, we perform 20 extra iterations after the system stabilizes to obtain operational KPIs.

\begin{figure}[H]
	\centering
		\includegraphics[height=7cm,keepaspectratio]{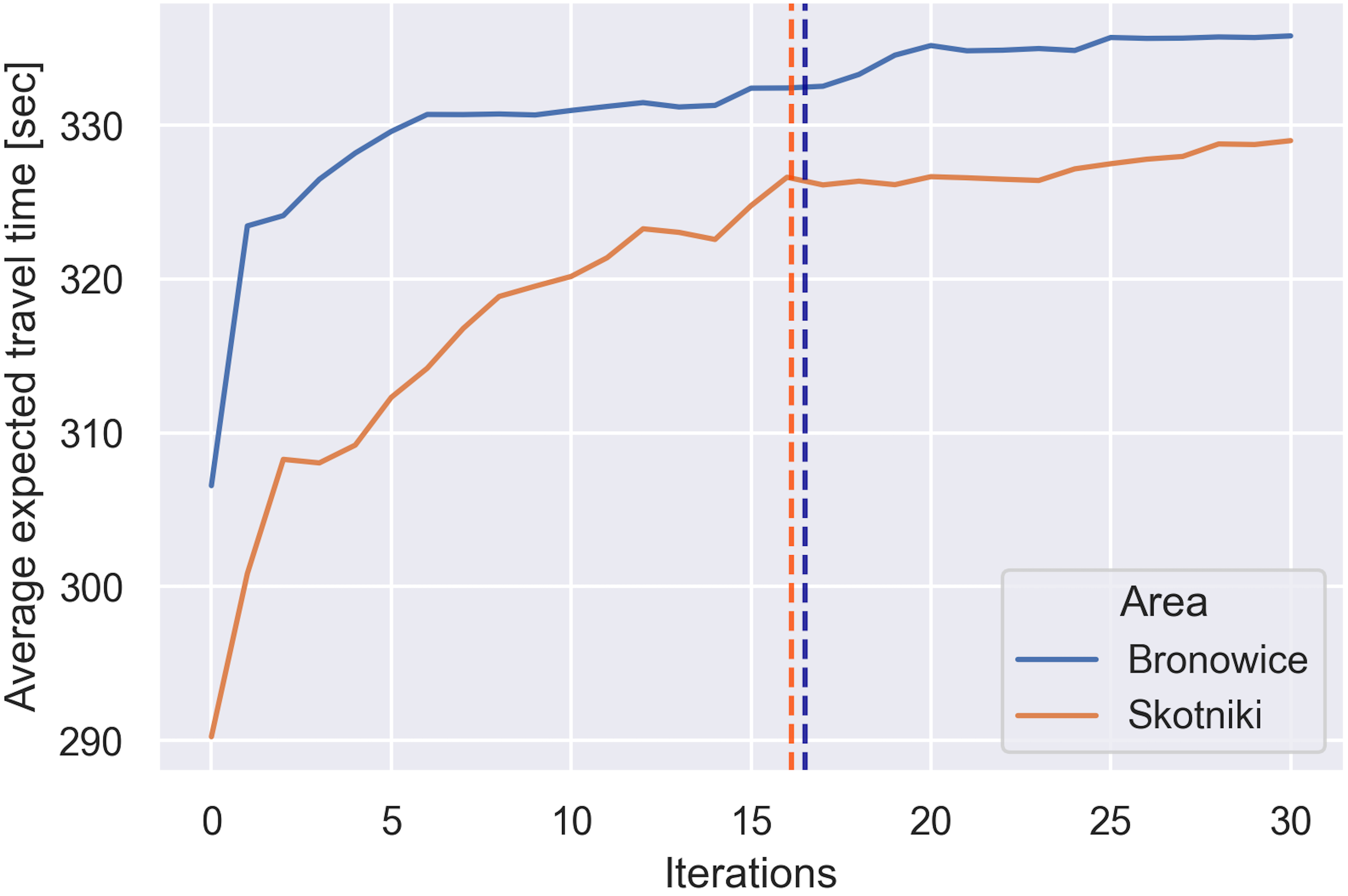}
	\caption{Stabilization of the average expected travel times through MSA over 30 iterations for Bronowice and Skotniki areas.
    The initial travel times at the first iteration correspond to raw travel times estimated for solo rides, before any learning or adaptation occurs.
    Travel times initially vary as travelers adapt to the new feeder bus system, but gradually stabilize as the system converges.
    Both areas show a similar trend of initial variation followed by convergence, though the specific travel times and rates of convergence differ. 
    The dashed lines represent the mean value of the convergence iteration, for Bronowice in blue and for Skotniki in orange, indicating the average convergence iteration around 16 (the mean over 100 demand replications).}
    \label{fig:msa_iterations}
\end{figure}

Figure \ref{fig:msa_iterations} illustrates a representative simulation run, depicting the stabilization of average expected travel times through MSA iterations for Bronowice and Skotniki areas. The figure provides a visual representation of the learning process for a single demand replication, showing how expected travel times stabilize over the iterations. 

As depicted in Figure \ref{fig:msa_iterations}, the early iterations are characterized by an adaptation phase, where expected travel times initially fluctuate before stabilizing. This reflects the dynamic shifts in demand as travelers refine their choices and adjust to the new feeder bus system. 
The expected travel times at the first iteration represent the baseline, calculated for solo rides prior to the learning process.
Assuming each iteration represents a day of operation, the convergence of travel times shows that traveler behavior stabilizes after approximately two weeks of simulated operation in our MSA model.
This observation emphasizes the importance of considering learning effects and system adaptation when implementing new transport services.

% Understanding this stabilization process ensures a more accurate evaluation of the feeder system performance, enhancing further analysis of traveler decision-making and service efficiency.

\subsection{Feeder choice probabilities}
\label{subsec:results_demand_distrib}

%Multiple replications of various demand configurations are essential for capturing the stochastic nature of travel behavior and ensuring statistically significant results for KPIs. SimFLEX enables multiple replications of demand scenarios for selected urban areas, facilitating the generation of diverse traveler choices. 
For a detailed understanding of individual decision-making, we analyze a single run of the method before aggregating results over multiple replications. For this, we generate a demand scenario, compute utilities and feeder choice probabilities for each traveler, and then sample actual feeder users based on these probabilities. Histogram of feeder probabilities illustrates how likely different travelers are to choose the feeder service under the given conditions.

\begin{figure}[H]
	\centering
		\includegraphics[height=7cm,keepaspectratio]{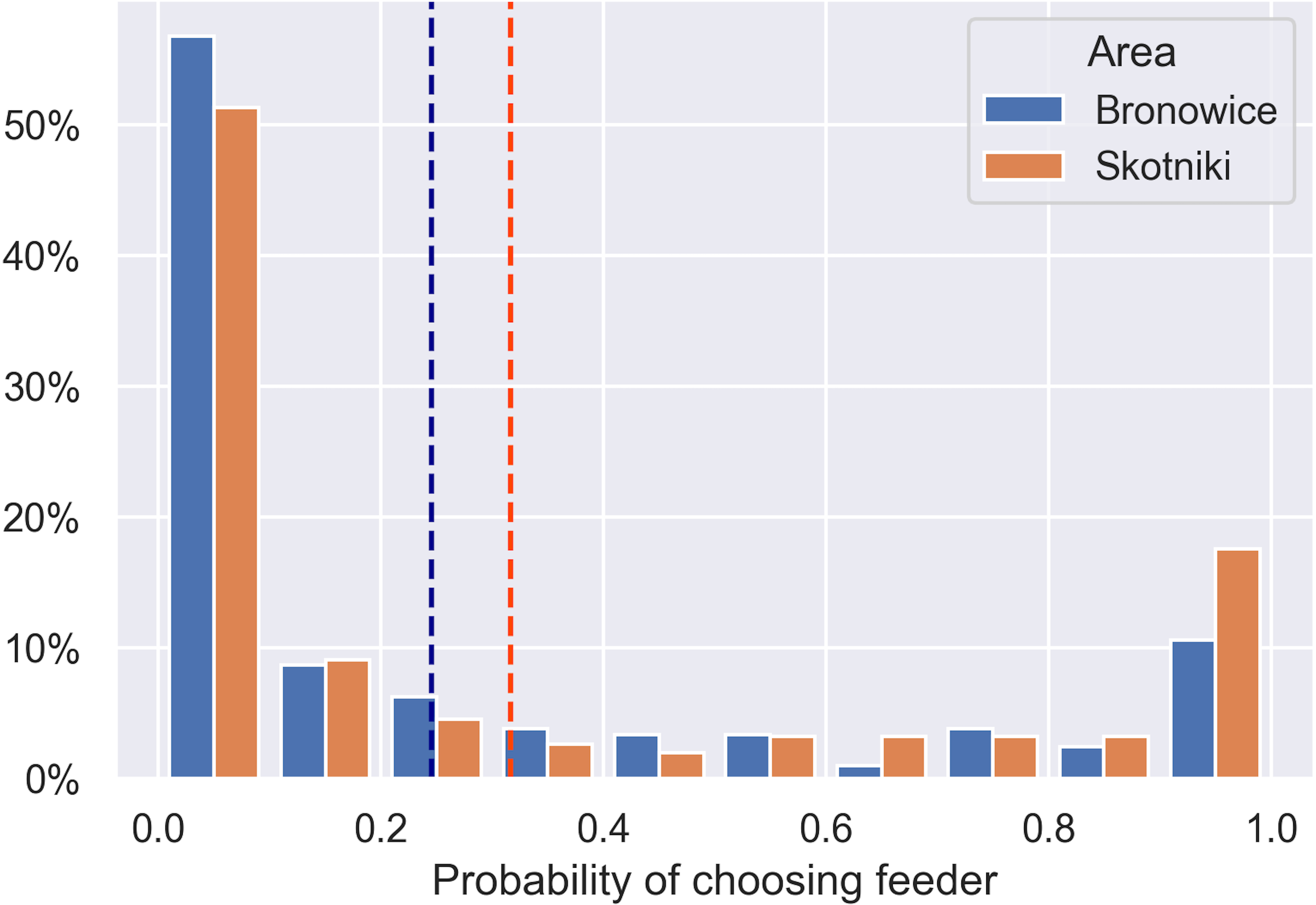}
	\caption{The probability distribution of feeder bus service choice in Bronowice and Skotniki for a single demand replication. 
    The figure shows that in Bronowice, slightly more than half of travelers (57\%) and in Skotniki roughly half (51\%) have almost 0\%  probability of choosing the feeder. Meanwhile, a tenth of travelers (10\%) in Bronowice and approximately a sixth (17\%) in Skotniki have a 100\% probability of choosing the feeder. The mean probability of feeder choice is 0.23 for Bronowice and 0.35 for Skotniki, as indicated by the dashed lines. These values highlight differences in the feeder service adoption between the two areas in this single demand scenario. 
    % Before combining results from a series of replications, observed differences demonstrate local variations in service attractiveness and traveler preferences in one method replication.
    }
    \label{fig:prob_feeder}
\end{figure}

Figure \ref{fig:prob_feeder} presents the probability distribution of choosing the feeder bus integrated with the public transport system for a single demand replication for analyzed areas, Bronowice and Skotniki. In both regions, a substantial portion of travelers exhibit a low probability of choosing the feeder service, with approximately 57\% in Bronowice and 51\% in Skotniki, indicating that in many cases, no travelers opted for the feeder bus service on the analyzed replication.
On the other hand, a notable fraction of travelers (around 17\% in Skotniki and 10\% in Bronowice) show a high probability (close to 1) of choosing the feeder service, with a slightly greater prevalence in Skotniki. The remaining probability values are distributed across intermediate ranges, with Skotniki displaying a slightly higher frequency of probabilities above 0.5. Additionally, the mean probability of the feeder choice, marked by the dashed lines, is lower in Bronowice (around 0.23) compared to Skotniki (nearly 0.35), suggesting that, on average, the feeder service is a more attractive option for travelers in Skotniki.

Analyzing a single demand replication for Bronowice and Skotniki, representing one day's traveler choices, allows for a direct comparison of feeder service prioritization in these areas, showing local differences in service attractiveness, before making more general conclusions from multiple runs of the algorithm.

% The results indicate that while the feeder service is viable in both locations, it is favored more consistently in Skotniki than in Bronowice. 
% However, to account for demand variability and achieve a more reliable evaluation of feeder service performance, we perform a series of demand replications. This approach allows to generalize the findings, capture uncertainties in traveler choices, and compare the potential of different areas based on overall service effectiveness.

\subsection{Comparative analysis of study areas}
\label{subsec:results_area_comparison}

%SimFLEX is primarily designed to calculate performance metrics for feeder bus services, with its main practical application being the comparison of urban areas to determine the most suitable location for the service implementation.
To assess the feasibility and relative attractiveness of feeder bus services in Bronowice and Skotniki, and to conduct a comparative analysis of these areas, we perform simulations, where we generate demand, evaluate user choices and model  the user learning process using the MSA, as outlined in Section \ref{subsec:results_config}. After system stabilization, we calculate KPIs, including effectiveness metrics of the integrated feeder system (detailed in Section \ref{subsubsec:results_main_KPI}) and ExMAS-derived operational KPIs of the feeder buses (detailed in Section \ref{subsubsec:results_exmas_KPI}).

\subsubsection{Effectiveness metrics of the integrated feeder system}
\label{subsubsec:results_main_KPI}

\begin{figure}[H]
	\centering
    \includegraphics[width=\textwidth,height=10cm,keepaspectratio]{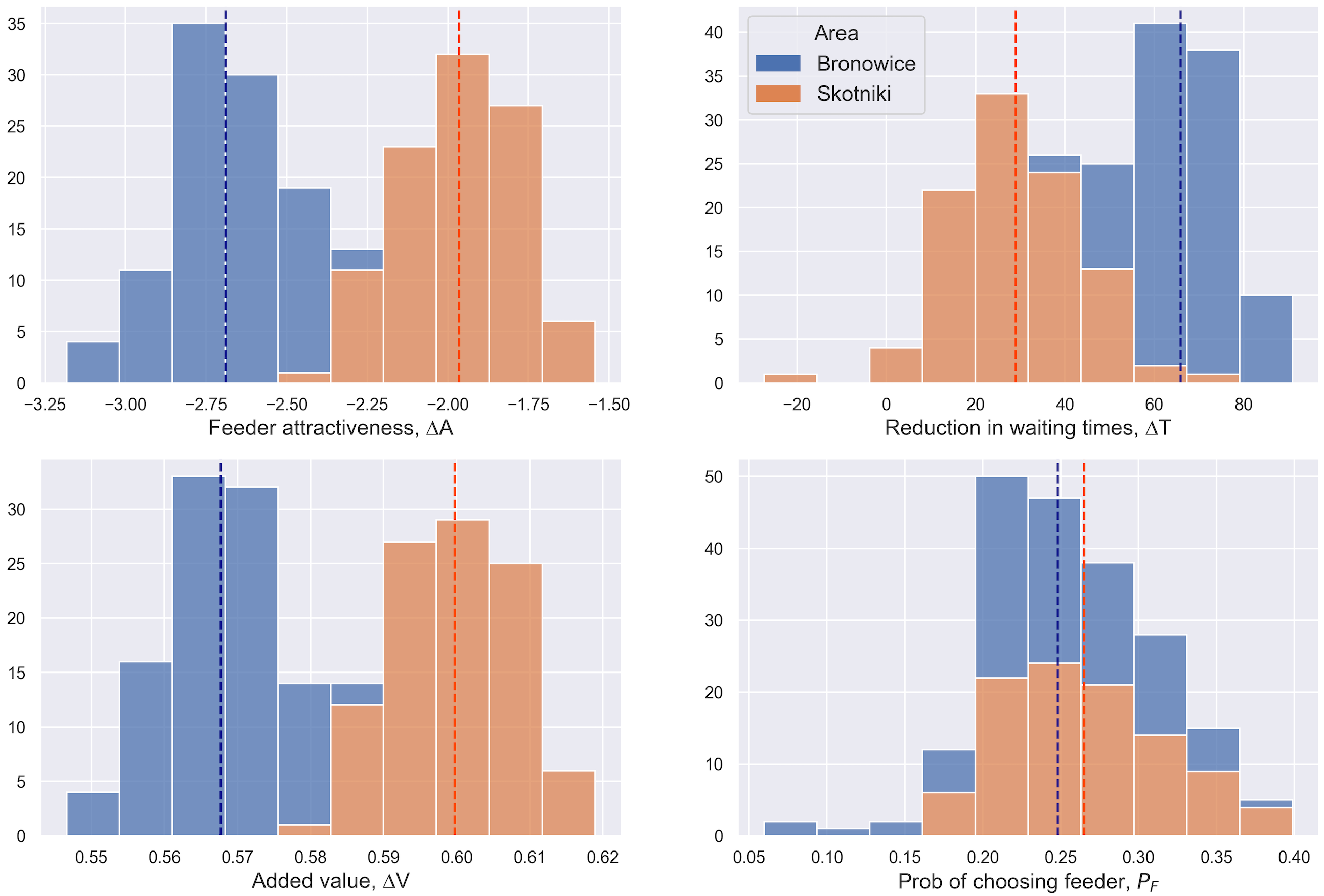}
	\caption{The distribution of integrated feeder system KPIs across the 100 demand replications with their mean values (illustrated with dashed lines) for Bronowice and Skotniki areas.
    Skotniki generally exhibits higher (less negative) values for feeder attractiveness (\(\Delta A\)), suggesting it is perceived as more attractive compared to Bronowice. In terms of waiting time reduction (\(\Delta T\)), Bronowice shows a distribution concentrated at higher positive values, indicating a more significant reduction in waiting times. For added value (\(\Delta V\)) Skotniki has a higher mean value and a narrower data spread, implying a slightly better added value compared to Bronowice, which shows a similar spread but with slightly lower mean values.
    Finally, the histograms of the probability of choosing the feeder (\(P_F\)) indicate that Skotniki has a distribution skewed towards higher probabilities, suggesting a greater likelihood of choosing the feeder service compared to Bronowice.
    }
    \label{fig:main_kpis}
\end{figure}

The primary metrics that describe the effectiveness of the feeder system integrated with public transport, used for Bronowice and Skotniki comparison (e.g., the probability of choosing feeders (\(P_{F}\), Eq. \ref{eq:P_F}), feeder attractiveness (\(\Delta A\), Eq. \ref{eq:Delta_A}), reduction in waiting time (\(\Delta T\), Eq. \ref{eq:Delta_T}), overall added value (\(\Delta V\), Eq. \ref{eq:Delta_V}) along with their respective variances) are summarized in Table \ref{table:main_kpis}, while their distributions are depicted in Fig. \ref{fig:main_kpis}. These indicators are the mean values of 100 demand replications, calculated after the system's stabilization (30 MSA iterations and an additional 20 iterations).

\begin{table}[!ht]
\caption{Evaluation of feeder service potential: Bronowice and Skotniki feeder system-level indicators comparison}
\label{table:main_kpis}
\vspace{0.3cm}
\centering
\resizebox{\textwidth}{!}{
    \begin{tabular}{ | c | c | c | c | c | c | c | c | c |}
        \hline
        \textbf{Area} & $ \boldsymbol{P_F} $ & $ \boldsymbol{Var(P_F)} $ & $ \boldsymbol{\Delta A} $ & $ \boldsymbol{Var (\Delta A)} $ & $ \boldsymbol{\Delta T} $ & $ \boldsymbol{Var (\Delta T)} $ & $ \boldsymbol{\Delta V}$ & $ \boldsymbol{Var (\Delta V)}$ \\
        \hline
        \multicolumn{1}{|l|}{Bronowice} & 0.248          & 0.003 & -2.690           & 0.034 & \textbf{65.904}  & 107.691 & 0.568          & 0.0001 \\
        \multicolumn{1}{|l|}{Skotniki}  & \textbf{0.265} & 0.003 & \textbf{-1.966}  & 0.036 & 28.925           & 224.955 & \textbf{0.600} & 0.0001 \\
        \hline
    \end{tabular}
    }
\end{table}

The comparison reveals that Skotniki presents advantages across the majority of key metrics.
Skotniki exhibits a marginally higher probability of passengers choosing feeder services (\(P_{F}\) = 0.265 compared to Bronowice \(P_{F}\) = 0.248). Furthermore, Skotniki demonstrates improved feeder attractiveness (\(\Delta A\) = -1.966), indicating enhanced accessibility compared to Bronowice (\(\Delta A\) = -2.690). Additionally, the overall added value of the feeder service is higher in Skotniki (\(\Delta V\) = 0.600) than in Bronowice (\(\Delta V\) = 0.568), suggesting improved connectivity and overall user satisfaction in the Skotniki area. Conversely, Bronowice offers a greater reduction in waiting time (\(\Delta T\) = 65.904), exceeding the time savings observed in Skotniki (\(\Delta T\) = 28.925). %This highlights a notable advantage for Bronowice in terms of time efficiency.
While most metrics demonstrate consistent results across replications, with relatively small variances, \(\Delta T\) shows significant variability. This requires further investigation to identify the contributing factors.

The distributions of the feeder system indicators across demand replications reveal that Skotniki generally exhibits higher values for most indicators, suggesting potentially superior performance. Specifically, Skotniki shows higher distributions and mean values for feeder attractiveness, added value, and probability of choosing the feeder service. 
Skotniki consistently exhibits higher (less negative) feeder attractiveness values across its distribution, with most values centered between -2.0 and -2.5, while Bronowice's distribution is shifted towards more negative, with most values centered between -2.5 and -3.0, indicating a lower perceived attractiveness.
In contrast, Bronowice's distribution of reduction in waiting time is concentrated at significantly higher values, primarily in the range of 60 - 80 seconds, demonstrating the advantage in reducing waiting times. Skotniki's distribution is clustered at lower values, mostly in the range of 20 - 40 seconds, indicating a less substantial reduction in waiting times. 
Furthermore, the distribution of added value shows that Skotniki generally exhibits slightly higher values, mostly around 0.6, compared to Bronowice, which has values centered around 0.55. Similarly, the distribution of the probability of choosing feeder shows that Skotniki's distribution is shifted slightly towards higher probabilities, in the range 0.25 - 0.3, compared to Bronowice, which is in the range 0.2 - 0.25, suggesting a potentially higher likelihood of passengers choosing the feeder service in Skotniki.

\subsubsection{Operational KPIs of the feeder bus service}
\label{subsubsec:results_exmas_KPI}

\begin{figure}[!ht]
	\centering
		\includegraphics[width=\textwidth,height=\textheight, keepaspectratio, scale=0.6]{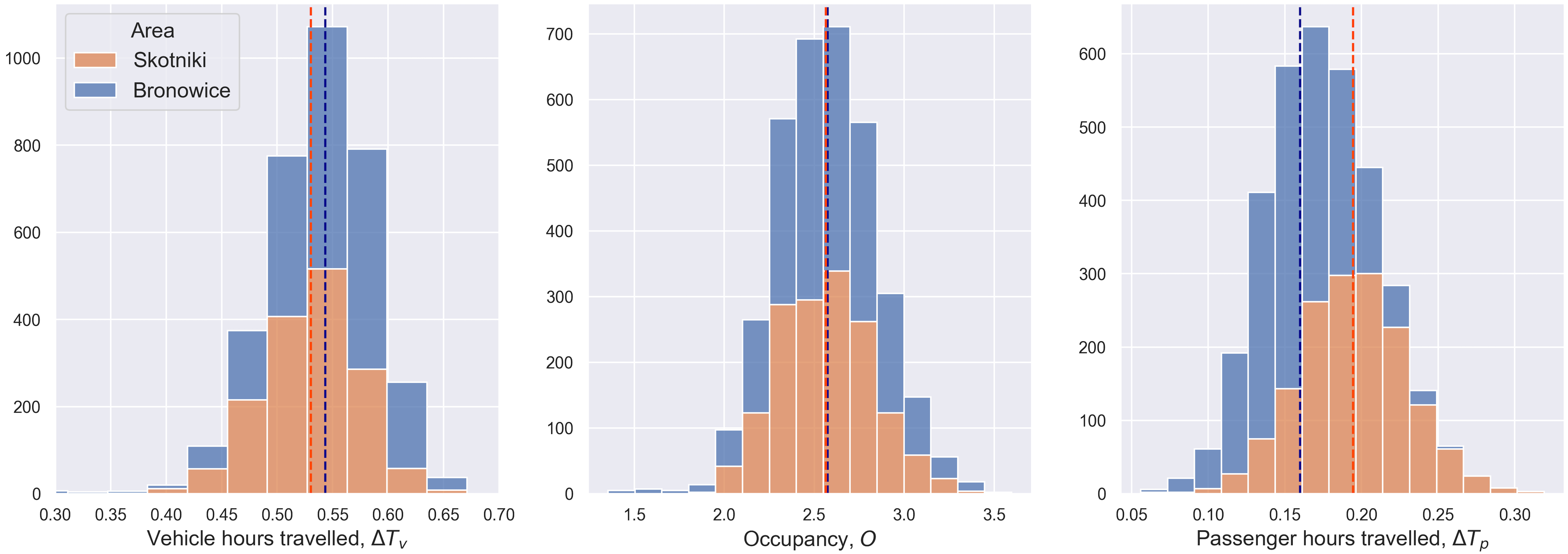}
	\caption{Distributions of relative changes in KPIs (with mean values depicted by dashed lines) following the implementation of feeder bus services compared to solo rides in Bronowice and Skotniki.  
    For vehicle-hours traveled (\(\Delta T_v\)), Bronowice displays a distribution with a higher mean increase, indicating a greater impact on vehicle operation. Similarly, Bronowice vehicle occupancy (\(O)\) is higher on average in Bronowice, suggesting better utilization of vehicle capacity. Conversely, for passenger-hours traveled (\(\Delta T_p)\), Skotniki shows a distribution with a higher mean increase, indicating a greater impact on overall passenger travel time despite the similar distribution shapes. These differences highlight the varying effects of the feeder system across the two areas, with Bronowice experiencing improved vehicle utilization and Skotniki showing a larger increase in passenger travel time.
    }
    \label{fig:ExMAS_kpis}
\end{figure}

For our comparative analysis, we also investigate the operational performance of the feeder buses across the analyzed areas, Bronowice and Skotniki, by using KPIs derived with the ExMAS algorithm. 
Based on simulation results obtained under the configuration described in Section \ref{subsec:results_config}, the histograms (Fig. \ref{fig:ExMAS_kpis}) depict the distributions of relative KPI changes for the feeder bus service in comparison to solo ride baseline (calculated based on \citet{Shulika:2024}).

We begin by analyzing the distribution of \(\Delta T_v\), which represents the change in vehicle-hours traveled. 
The distributions for both areas are relatively similar, with the majority of values concentrated between 0.5 and 0.6. The mean \(\Delta T_v\) is approximately 0.55 for Bronowice and 0.52 for Skotniki, indicating a slightly greater increase in vehicle-hours in Bronowice. Both distributions have minimal tails, indicating that changes in vehicle-hours are consistent across trips in both areas. Overall, the introduction of feeder services results in a comparable decrease in vehicle-hours in both regions, with a marginally stronger effect observed in Bronowice.

Next, we examine the distribution of vehicle occupancy (\(O\)), defined as the average number of passengers per vehicle-hour. The peak occupancy for Bronowice is observed between 2.4 and 2.6, with a mean value of approximately 2.5, whereas Skotniki has a slightly lower mean of 2.4. The overall distribution for the Bronowice area is shifted slightly to the right, indicating better utilization of vehicle capacity.

Finally, we analyze the distribution of \(\Delta T_p\), representing the additional total travel time experienced by feeder bus travelers compared to solo ride scenario. The values are primarily concentrated between 0.15 and 0.25, with a mean of approximately 0.17 for Bronowice and 0.19 for Skotniki, indicating that passengers in Skotniki experience a slightly greater increase in travel time on average. Overall, while both areas experience increased passenger-hours with the introduction of feeders, the impact is more pronounced in Skotniki.

% ExMAS KPIs

\subsection{Sensitivity analysis of KPIs to ASC variations}
\label{subsec:results_asc_analysis}

A significant advantage of the SimFLEX methodology is that it enables the analysis of the sensitivity of outputs to variations in alternative-specific constant. ASCs represent the average effect of unobserved factors on mode choice utility, and they are typically unknown a priori for new service implementations. 

To conduct the sensitivity analysis, we perform simulations with five area sample replications for each ASC value (\(N = 5\)). We vary the ASC parameter across a range from 0 to 5, with 50 evenly spaced points sampled from a uniform distribution within this interval. This approach allows for the exploration of a wide range of potential user preferences and assessing the stability and reliability of method outputs.

\begin{figure}[!h]
	\centering
		\includegraphics[width=\textwidth,height=10cm,keepaspectratio]{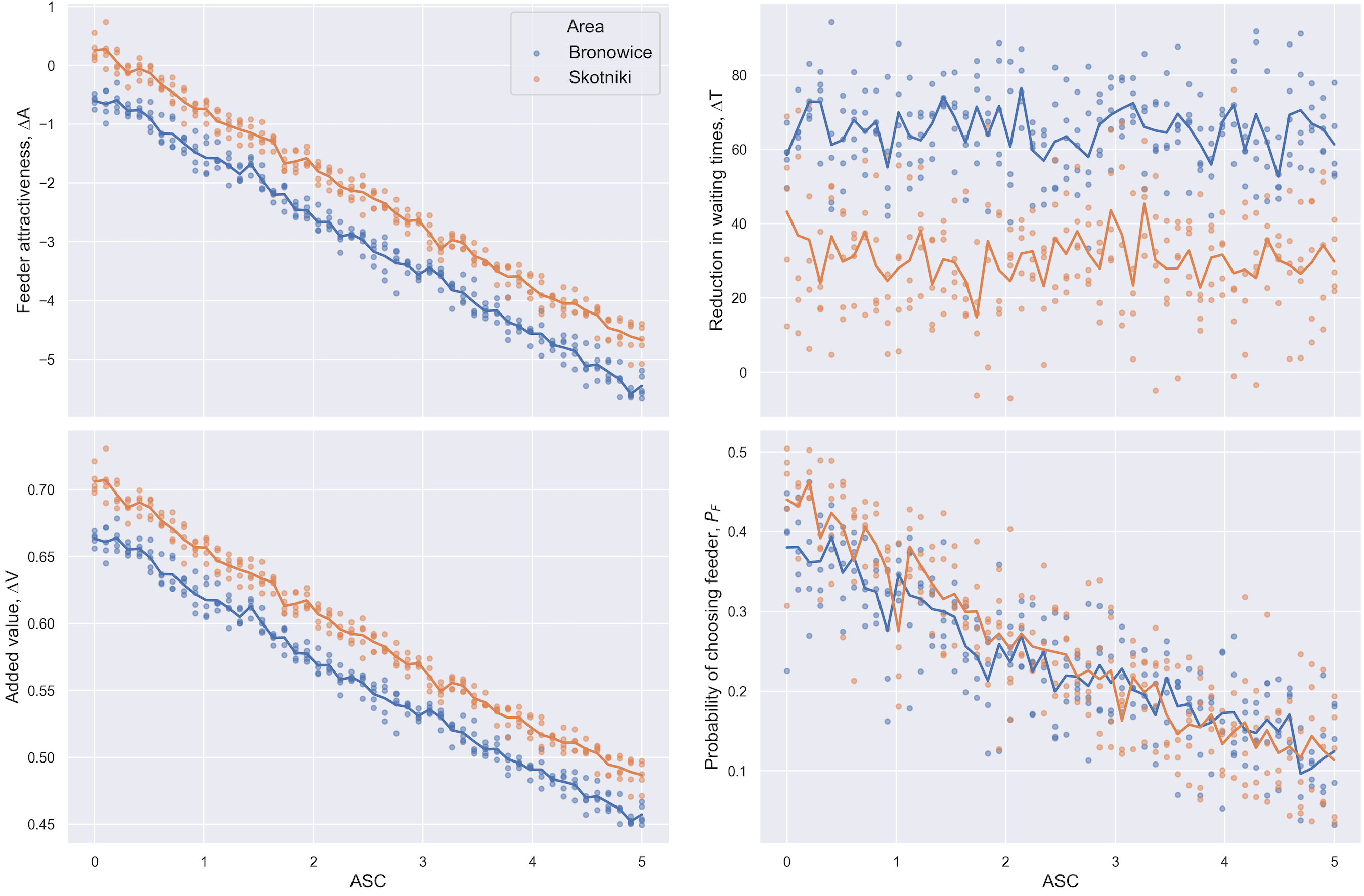}
	\caption{
    The results remain consitent when the behaviural parameter ($ASC$) varies. The impact of ASC variations (ranging from 0 to 5) on four key metrics: feeder attractiveness (\(\Delta A\)), reduction in public transport waiting times (\(\Delta T)\), overall added value (\(\Delta V)\), and the probability of choosing the feeder service (\(P_F\)). Each data point represents the average value of the respective metric across five replications per ASC value, with solid lines indicating trends for each area.
    %\\ The analysis compares two areas in Krakow, Bronowice and Skotniki, highlighting differences in sensitivity to ASC changes. Skotniki generally shows higher feeder attractiveness and added value, particularly at lower ASC values, while Bronowice exhibits a greater reduction in public transport waiting times. The probability of choosing the feeder service decreases with increasing ASC for both areas, indicating a reduced preference for feeder services as ASC increases. Waiting time reductions remain relatively stable across ASC values, suggesting that this metric is less influenced by user preference shifts and more dependent on network efficiency.
    }
    \label{fig:asc_vs_kpis}
\end{figure}

The results of this analysis, presented in Fig. \ref{fig:asc_vs_kpis}, reveal distinct patterns in how the choice of ASC affects the primary KPIs across the two analyzed areas: Bronowice and Skotniki. Notably, Skotniki consistently demonstrates greater potential across most indicators, particularly in feeder service attractiveness (\(\Delta A\))  and added value (\(\Delta V)\), suggesting enhanced feeder effectiveness in this area. The results also highlight a general trend of decreasing feeder attractiveness, added value, and probability of choosing the feeder service (\(P_F\)) as the ASC increases. In contrast, the reduction in public transport waiting times (\(\Delta T)\) remains relatively stable across the range of ASC values.
% Furthermore, the probability of choosing the feeder service exhibits a consistent decline with increasing ASC values.

% Skotniki performs better on qualitative and utility-focused metrics, such as feeder attractiveness and added value, making it more favorable at low to moderate ASC levels. In contrast, Bronowice excels in reducing public transport waiting times, potentially making it more attractive to users prioritizing time efficiency or relying heavily on public transport.
% General trend across all indicators suggests that indicators decrease as ASC increases, except for the reduction in waiting times for PT, which remains relatively stable. This could be explained by the fact that ASC influences utility-focused metrics, such as feeder attractiveness, probability of choosing a feeder, and added value, leading to their decline. In contrast, the reduction in waiting times for PT is likely more dependent on the structural characteristics of the public transport network and scheduling efficiency rather than ASC, resulting in its relative stability.

% \newpage
\clearpage

\section{Discussion and Conclusion}
\label{sec:discussion}

This paper introduces SimFLEX, a methodology developed to empower stakeholders in pre-deployment evaluation of the feasibility and performance of on-demand feeder bus services. A key feature of SimFLEX is its ability to compute service performance metrics under conditions of unknown demand and user perception through microsimulations, enabling comparative analysis of candidate urban areas and facilitating the selection of the optimal site for service deployment.
The framework enables evaluating both, the operational efficiency of feeder buses and the overall performance of the integrated feeder-public transport system.
System-level KPIs include feeder attractiveness, added value, and waiting time reductions, while operational indicators comprise vehicle-hours, passenger-hours traveled, and vehicle occupancy.
These metrics capture changes in transportation system performance due to the introduction of on-demand feeder bus services, allowing conclusions to be drawn about which area would benefit more from feeder deployment and support long-term successful service use.
% SimFLEX provides a comprehensive approach for evaluating new services before implementation. Through multiple replications, the framework enables estimating travel demand using available macro-level data and modeling traveler mode choices based on perceived value (utilities) without relying on pre-existing user perception data. 
To achieve this, we integrate the following components into our methodology: ExMAS for optimizing ride-pooling operations, MSA for simulating traveler learning and achieving system stabilization, and OTP for integrating the feeder service with the public transport system.

To demonstrate the practical applicability of SimFLEX, we present a case study analyzing the effects of introducing on-demand feeder bus services in two distinct urban areas in Krakow, Poland: Bronowice and Skotniki. This analysis illustrates how SimFLEX can be used to assess and compare the potential impacts of feeder services in different urban environments, providing decision-makers with information on which area is more suitable for feeder service deployment.

We analyze the convergence of average expected travel times in both areas over multiple MSA iterations, simulating the traveler learning process (Fig. \ref{fig:msa_iterations}). The results indicate that the system gradually stabilizes (in average, during a two-week period) as travelers adjust their choices, which is essential for accurately calculating KPIs and assessing the long-term performance of the integrated transportation system.

To account for demand variations and ensure system stabilization, the main KPIs are estimated after multiple algorithm runs. Each run simulates the learning process over MSA iterations for a specific demand replication. We examine a single demand replication in detail to visualize the distribution of feeder choice probabilities (Fig. \ref{fig:prob_feeder}) and highlight potential variations in user behavior that may be hidden by aggregated results.
The analysis of feeder choice probabilities in a single demand replication shows that feeder selection is generally more favorable in Skotniki than in Bronowice, suggesting that the feeder service is a more attractive option for travelers in Skotniki. This difference could be attributed to variations in public transport accessibility, planned feeder service characteristics, or local demand patterns. The lower mean probabilities in Bronowice suggest that travelers either have better alternative transport options or that the feeder service is less competitive compared to other modes.
%Overall, the results suggest that feeder bus integration is more effective in Skotniki, where the probability distribution is more balanced and includes a greater tendency for feeder selection. The higher proportion of zero-probability observations in Bronowice suggests potential challenges in feeder adoption. 
% However, further analysis is needed to confirm these assumptions and to determine the specific factors influencing traveler choices.

A quantitative comparison of integrated feeder transit system KPIs between Bronowice and Skotniki reveals significant differences in service potential (Table \ref{table:main_kpis}). Skotniki exhibits a higher average probability of feeder service selection (0.265 compared to 0.248 in Bronowice) and a greater (approximately by 30\%) feeder service attractiveness (-1.966 compared to -2.690 in Bronowice). Furthermore, Skotniki demonstrates a higher (up to 7\%) added value (0.600 compared to 0.568 in Bronowice). In contrast, Bronowice shows a substantially greater potential for reducing traveler waiting times (nearly by 77\%), with an average reduction of 65.904 seconds compared to 28.925 seconds in Skotniki. These numerical differences support the data from the figure (Fig. \ref{fig:main_kpis}), highlighting the distinct advantages and challenges of feeder service implementation in each area.

The analysis of KPIs reveals distinct patterns in the potential for on-demand feeder bus service implementation in Bronowice and Skotniki (Fig. \ref{fig:main_kpis}). Skotniki demonstrates a higher average probability of users choosing the feeder service, coupled with a greater service attractiveness and added value. 
Conversely, Bronowice exhibits a significant potential for reducing traveler waiting times for public transport. However, the lower average probability of feeder service selection and the distribution of choice probabilities suggest that the service may face challenges in attracting users. The concentration of zero-probability values in Bronowice indicates that a substantial portion of the population may consistently opt for alternative modes of transport.
Based on these findings, Skotniki appears to be a more suitable area for the feeder service implementation. The higher average probability of feeder selection, greater attractiveness, and added value suggest a greater likelihood of successful integration and user adoption of feeders in this area.

Based on the analysis of operational KPIs, the implementation of feeder bus services in Bronowice and Skotniki reveals distinct impacts. Bronowice exhibits a higher average vehicle occupancy and increased vehicle-hours traveled. However, Skotniki shows a larger increase in passenger-hours traveled, suggesting potentially greater delays for travelers. Therefore, while Bronowice achieves better vehicle utilization, Skotniki results in an increased time burden for travelers. In terms of operational efficiency, neither area clearly outperforms the other, as each demonstrates a trade-off between vehicle utilization and passenger travel time.

To further assess the stability and reliability of our method outputs, we conduct a sensitivity analysis across a range of ASC values. The results consistently confirm Skotniki as the superior candidate for feeder service implementation across all ASC variations (Fig. \ref{fig:asc_vs_kpis}). This consistency strengthens our findings that Skotniki's advantages result from differences in travel demand and service characteristics, not specific ASC assumptions.

However, the decision on whether Bronowice or Skotniki is a more suitable area for the feeder service implementation depends on the priorities of the transportation authorities and the specific goals of the service deployment. If minimizing travel time is the primary objective, Bronowice appears to be the preferable choice. However, this choice may cause lower overall service adoption and potential long-term benefits due to lower service attractiveness. 
Conversely, if increasing accessibility and consistent service usage are prioritized, Skotniki is a more appropriate choice. The higher feeder service attractiveness and the added value in Skotniki suggest a greater likelihood of consistent ridership and integration with the existing transportation system.

\subsection{Limitations and future research}
\label{subsec:limitations}

Despite the limitations, SimFLEX opens avenue for the new, more detailed and realistic assessment for newly introduced feeded services. Allowing series of extenstions, such as:

\begin{enumerate}
\tightlist
    \item Analysis of the impact of travellers' heterogeneity on simulation outcomes.
    Investigate how variance of travel behaviour among travellers (e.g. value of time or $ASC$) affects simulation results. 
    
    \item Determination of the optimal area.
    This, challenging computationally problem becomes now feasible, where SimFLEX can be used to determine the optimal service area.
    %Examine the impact of feeder service area size and geometric configurations on system efficiency. This includes identifying the trade-offs between operational costs and service accessibility to identify the most effective design parameters.

    \item Detailed socio-demgraphics and trip characteristics.
    To include varying propensity to use feeder service depending on age, trip-purpose, urgency, etc.

    \item Expansion of the scope of the study to a broader range of urban areas.
    Extend the evaluation framework to include different geographical regions, urban and suburban environments, and varied demographic compositions to generalize findings and improve the reliability of results.

    \item Integration of alternative travel modes.
    Incorporate additional transportation options such as private car usage, ride-sharing, or micromobility solutions to assess the relative benefits of feeder transit systems in multimodal environments.

    \item Incorporation of advanced learning models beyond MSA.
    Develop and integrate more sophisticated learning models to improve service planning. Potential approaches include 
    moving averages, exponential smoothing, Kalman filters, reinforcement learning, deep neural networks, and Bayesian inference techniques.
\end{enumerate}

SimFLEX is a comprehensive and adaptable simulation framework that enables decision-makers to evaluate urban areas based on multiple indicators, helping prioritize locations with the highest potential for successful feeder service deployment.
By analyzing both system-wide and feeder-specific KPIs, it provides a comprehensive assessment of integrated feeder transit system, while the inclusion of sensitivity analysis ensures the stability and reliability of method outputs.

A key strength of SimFLEX is its emphasis on user perception modeling before service deployment, increasing the likelihood of a service acceptance and success in the targeted area. 
Moreover, its flexibility allows application across various urban environments, adapting different transportation infrastructures and demand patterns.
By addressing the absence of systematic, location-specific evaluation methods for planned feeder services, SimFLEX serves as a novel decision-support tool for urban mobility planning. Its ability to estimate travel demand with limited data, combined with its focus on optimizing both operational efficiency and user experience, makes SimFLEX a valuable asset for future urban transportation projects.

\section{CRediT authorship contribution statement}

\textbf{Hanna Vasiutina}: Investigation, Methodology, Software, Writing – review \& editing, Writing – original draft. 
\textbf{Olha Shulika}: Data curation, Investigation, Formal analysis, Software, Writing – review \& editing, Writing – original draft. 
\textbf{Michał Bujak}: Data curation, Investigation, Formal analysis, Software, Writing – review \& editing.
\textbf{Farnoud Ghasemi}: Data curation, Investigation, Formal analysis, Software, Writing – review \& editing.
\textbf{Rafał Kucharski}: Conceptualization, Methodology, Supervision, Writing – review \& editing, Writing – original draft.

\section{Declaration of competing interest}
The authors declare that they have no known competing financial interests or personal relationships that could have appeared to influence the work reported in this paper.

\section{Code and data availability}
\label{sec:github}

The simulation code and input data are available in the repository at \url{https://github.com/anniutina/SimFLEX}. Due to GDPR restrictions, the population distribution and ODM are not publicly accessible. The simulation results presented in this paper are also stored in the same repository.

\section{Acknowledgements}
This research was co-funded by the European Union’s Horizon Europe Innovation Action under grant agreement No. 101103646.

\section{Declaration of generative AI and AI-assisted technologies in the writing process}

During the preparation of this work the authors used ChatGPT (by OpenAI) in order to improve the readability and language of the manuscript. After using this service, the authors reviewed and edited the content as needed and take full responsibility for the content of the published article.

%% Use \subsubsection, \paragraph, \subparagraph commands to 
%% start 3rd, 4th and 5th level sections.
%% Refer following link for more details.
%% https://en.wikibooks.org/wiki/LaTeX/Document_Structure#Sectioning_commands

%\subsubsection{Subsubsection}
%% Inline mathematics is tagged between $ symbols.
%This is an example for the symbol $\alpha$ tagged as inline mathematics.

%% For citations use: 
%%       \cite{<label>} ==> [1]

%%
%% If you have bib database file and want bibtex to generate the
%% bibitems, please use
%%
%%  \bibliographystyle{elsarticle-num} 
%%  \bibliography{<your bibdatabase>}

%% else use the following coding to input the bibitems directly in the
%% TeX file.

%% Refer following link for more details about bibliography and citations.
%% https://en.wikibooks.org/wiki/LaTeX/Bibliography_Management

% \bibliographystyle{plainnat}
\bibliographystyle{abbrvnat}
%\bibliography{refs}

%% For numbered reference style
%% \bibitem{label}
%% Text of bibliographic item

\end{document}